# Response to the paper: Theoretical understanding of evolutionary dynamics on inhomogeneous networks

Christopher Li

Department of Physics, Rice University

**Abstract**

As a co-author of the paper "Theoretical understanding of evolutionary dynamics on inhomogeneous networks" (the paper), I would like to express my disagreement with the conclusion of the paper. In this response, I present a thorough examination of the paper's assertions and methods. Although I may disagree with several practices in the research group, I will confine my discussion to the academic analysis of the paper in this response.

## 1 Introduction

Evolutionary dynamics on graphs is a relatively new area of research that has attracted considerable interest. The fixation probability and fixation time of birth-death processes on complete bipartite graphs, which encompass star graphs, have been well-established through rigorous investigation [1] [2] [3] [4] [5] [6].

The work in "Theoretical understanding of evolutionary dynamics on inhomogeneous networks" (the paper) [7] primarily consists of a repetition of Ref. [1]. In addition, it presents an approximate model and a novel interpretation, both of which I believe to be meaningless.

Side note: The "B" in author name "Christopher B Li" that appears in the paper was a result of a typographical error. I do not have a middle name. The "B" may not appear in future editions.

## 2 Detailed Analysis of the Paper's Methods and Claims

### 2.1 First-Passage Method

In the supporting information of the paper (Fig.S1), first-passage method was used to calculate the transition probabilities. However, this method was unnecessary and resulted in complicating a straightforward problem. To see this, I use $a_n$ as an example to demonstrate.

When the central node is mutated and there are (n-1) mutations in the "leaves", the rate at which the system evolves to the (n + 1) state (mutated central node plus n mutated cells in the leaves) is simply r × 1 (the rate at which the central node produces offspring) multiplied by $(N-n)/(N-1)$ (The probability that a normal cell in the leaves is chosen to die).

Similarly, all other transition rates can be obtained without utilizing the so-called first passage method.

While the transition rates presented in the paper are factually correct, the use of the first-passage method is absolutely unnecessary and can cause confusion among readers.

I raised this concern with the group, and it was also brought up by one of the referees of the paper. Despite these concerns, the final version of the paper still presented this confusing argument.

## 2.2 The Approximate Model

In the paper, an approximate model was introduced. However, as I mentioned, the fixation probability can be easily found by Monk's martingale method in Ref. [2].

Given that there is an exact solution to the fixation probability [2] [1], the approximate model lacks meaningful or relevant insights.

One may argue that the approximate model gives an important time scale, however, this is not the case. The time scale of the discrete-time birth-death process has been given in Ref. [6] and Ref. [5] (Be careful with the definition of "fixation time" when reading the papers).

One may argue that previous research gave the Nln(N) time scale by using simulation which may be treated as "inferior". However, the approximate model underestimates the fixation time, it should be considered even more inferior than the simulation result.

Another important point worth noting is that in the continuous time model presented in the paper, if the parameter r is sufficiently large, the fixation time approaches zero. In contrast, for the discrete time model, the fixation time approaches N, the population size. In my view, this subtlety is significant and should have been explicitly mentioned in the paper. While it may or may not challenge the claims made in the paper, it is an important aspect to consider.

According to the paper, the significance of the approximate model and the use of a continuous time model lies in their ability to facilitate new interpretations. However, in the following section, I will present arguments to counter this claim.

### 2.3 Claims in the main text

In the conclusion of the paper, it says "The presented theoretical method allowed us to better understand the microscopic origin of fixation amplification that is accompanied by significant increase in fixation times".

In my opinion, this claim is false. The idea of "pathways" described in the paper is merely an illusion. While the concept might be helpful for those who struggle with mathematics, it is presented in a hand-waving manner and should not be included as a formal conclusion in the paper.

Moreover, the same "pathway" argument can be easily constructed using discrete-time models as well.

### 2.4 Unreliability of the Monte Carlo Simulation Codes

The Monte Carlo simulation codes employed by the research group were found (by me) to contain several bugs, making the simulation results unreliable occasionally. Upon joining the group, I corrected the codes. However, it is important to note that I cannot personally guarantee the validity of all simulations.

## 3 Conclusion

As one of the co-authors of the paper, I hold the opinion that the results presented in the paper lack publishable quality. A significant portion of the paper merely reiterates what others have already accomplished. Moreover, the conclusion drawn is not scientifically sound and is based on hand-waving arguments.

While it is my opinion that this pattern of repeating existing work and presenting meaningless interpretations has been observed in other publications from this research group, for the purposes of this discussion, I will focus solely on this particular paper.

I grant permission to all other authors of the paper to republish the paper without including my name.

I grant permission to Anatoly Kolomeisky's research group to use my previous work on the so-called "L-star model" (usually called complete bipartite graphs) in their future publications without including my name.

I do not give my consent to have my name included in any of their future publications, whether as a co-author or in the acknowledgements.

## 4 Acknowledgements